\documentclass[twocolumn,twoside,slac_two]{revtex4}
\usepackage{graphicx}
\usepackage{fancyhdr}
\def\slashchar#1{\setbox0=\hbox{$#1$}           % set a box for #1
   \dimen0=\wd0                                 % and get its size
   \setbox1=\hbox{/} \dimen1=\wd1               % get size of /
   \ifdim\dimen0>\dimen1                        % #1 is bigger
      \rlap{\hbox to \dimen0{\hfil/\hfil}}      % so center / in box
      #1                                        % and print #1
   \else                                        % / is bigger
      \rlap{\hbox to \dimen1{\hfil$#1$\hfil}}   % so center #1
      /                                         % and print /
   \fi}                                         %
\def\etmiss{\slashchar{E}_T}			%
\def\htmiss{\slashchar{H}_T}			%
\def\ptmiss{\slashchar{p}_T}			%

\begin{document}

%Title of paper
\title{Search for Leptoquarks with the D0 detector}

% Repeat the \author .. \affiliation  etc. as needed
%
% \affiliation command applies to all authors since the last
% \affiliation command. The \affiliation command should follow the
% other information

%\author{S.A. Uzunyan\footnote{speaker}\footnote{On behalf of the D0 collaboration}} 
\author{Sergey A. Uzunyan} 
\affiliation{Department of Physics, Northern Illinois University, DeKalb, IL 60115, USA\\
On behalf of the D0 Collaboration}

\begin{abstract}
We report on D0 searches for leptoquarks (LQ) predicted in extended gauge theories and composite models to
explain the symmetry between quarks and leptons. 
Data samples obtained with the  D0 detector from $p\bar{p}$ collisions at a center-of-mass 
energy of 1.96~TeV corresponding  to  intergrated luminosities of 1--4~fb${^{-1}}$ were analyzed.
No evidence for the production of such particles were observed and lower limits on leptoquark masses are set. 
\end{abstract}
%
%\maketitle must follow title, authors, abstract
\maketitle
\thispagestyle{fancy}
% body of paper here - Use proper section commands
% References should be done using the \cite, \ref, and \label commands
% Put \label in argument of \section for cross-referencing
%\section{\label{}}
%%%%%%%%%%%%%%%%%%%%%%%%%%%%%%%%%%
\section{Introduction}
Current theories~\cite{lq_theory} suggest that leptoquarks would come in three different
generations corresponding to the three quark and lepton generations. 
Leptoquarks would have color, fractional electric charge, both lepton and baryon numbers, 
and could be scalars or vectors. At the Tevatron leptoquarks could appear in pairs through   
$q\bar{q}$ annihilation (dominates for $M_{LQ}~>~100$~GeV) or $gg$ fusion, \newline
\centerline{$p+\bar{p} \rightarrow LQ+\overline{LQ}+X$}
or through the associated lepton production \newline
\centerline{$p+\bar{p} \rightarrow LQ+\bar{l}+X$} with the contribution of the last one being small.
The pair production cross section for scalar leptoquarks only depends on the strong coupling
constant and on the leptoquark mass. The vector leptoquark pair production cross section
also depends on anomalous $LQ-gluon$ couplings $k_G$ and $\lambda_G$, and the experimental
constrains are generally given for three models: ``Minimal Coupling'' ($k_G=1$, $\lambda_G=0$),
``Yang Mills'' ($k_G=0$, $\lambda_G=0$), and ``Minus Minus'' ($k_G=-1$, $\lambda_G=-1$).
%To obtain limits on scalar leptoqurak masses the next-to-leading order (NLO) pair production
%cross sections were used; for vector leptoquarks mass limits were obtained using the leading
%order theory. 

Leptoquarks decay into a charged lepton and a quark with a branching fraction $\beta$
or into a neutrino and a quark with a branching fraction ($1-\beta$). Thus
frations of a leptoquark pairs into the $llqq$, $l\nu qq$ and $\nu\nu qq$ final states 
are $\beta^2$, $2\beta(1-\beta)$ and $(1-\beta)^2$ respectively.

This report presents a summary searches for leptoquark pair production in the data sets 
collected with the D0 detector~\cite{d0_upgrade} during Run~II (started March 2001) of
the Fermilab Tevatron Collider. 
%
%All analysis presented in this report studied the pair production of leptoquarks$. 
% 
%%%%%%%%%%%%%%%%%%%%%%%%%%%%%%%%%%
\section{Generation independent leptoquark search. }
A search explored the final state where both leptoquarks decay into a neutrino and 
quarks assuming $\beta = 0$: $LQ\overline{LQ} \rightarrow \nu\bar{\nu}jj$. 
The corresponding detector signature is two acoplnar jets accompanied
by missing energy. No selection specific for jet flavor were applied. 
Thus the search results applied to all three generations of leptoquarks. 
The analysis used 2.5~fb$^{-1}$ of D0 Run~II data.

Events were recorded using triggers requiring two acoplanar jets  
and large missing transverse energy, $\etmiss$, 
or large $\htmiss$, the vector sum of the jets transverse energy $\htmiss \equiv|\sum_{jets}\vec{p_T}|$. The two leading jets were required to be in the central region $|\eta| < 0.8$ of the D0 detector and have  transverse momenta greater then 35~GeV. 
The multijet QCD background was suppressed by requiring $\etmiss$ to be greater then 75~GeV 
and with cuts on angular correlations between jets and $\etmiss$ directions: 
the azimutal angle between $\etmiss$ and the first jet and the minimal and maximal 
angles between any jets and $\etmiss$.  
To suppress the dominant standard model (SM)  background from $W(\rightarrow\ l\nu)+jets$ events a veto on events 
containing an isolated electron or muon with $p_T > 10$~GeV was applied,
and events with an isolated track $p_T > 5$~GeV were also rejected. 
The two final cuts on $\etmiss$ and $H_T=\Sigma_{jets}~p_T$ were optimized for different 
signals by minimizing the expected upper limit on the cross section in the absence of signal  

\begin{figure}[ht]
\centering
\includegraphics[height=65mm,width=80mm]{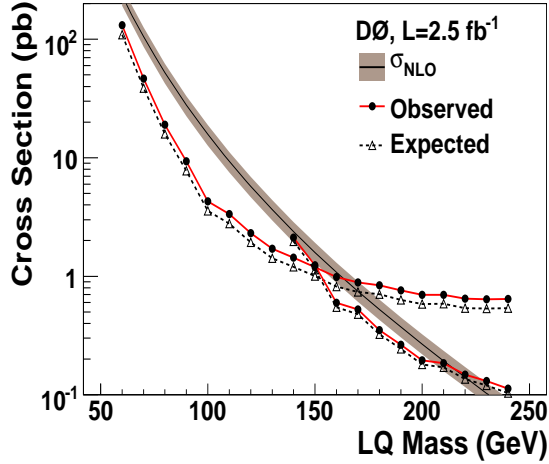}
\caption{Exclusion observed (circles) and expected (triangles) 95\% C.L. limits for the generation independent leptoquark search. The two graphs correspond to the low and high mass selections. The nominal NLO production cross section
is shown with shaded bands corresponding to the uncertainties due to choice of the parton distribution functions (PDFs) and variations of $\mu$ from  $0.5\times M_{LQ}$ to  $2 \times M_{LQ}$ added in quadrature.} 
\label{fig:lq_allgen}
\end{figure}

Table~\ref{tab:lq_allgen} shows the number of data, background and signal events 
after all selections for $M_{LQ} = 140$~GeV and $M_{LQ} = 200$~GeV leptoquark signals. 
No significant excess of data over the predicted backround was found. 
Figure~\ref{fig:lq_allgen} shows the observed and expected 95\% C.L. limits 
on scalar leptoquark pair production cross sections as a function of the LQ mass.
The observed and expected lower LQ mass limit of 214~GeV and 222~GeV respectively were obtained at the
intersection of the experimental cross section limits with the nominal theoretical production cross section 
calculated for the factorization and renormalization scale $\mu = M_{LQ}$. 
% Acconting uncentanties due to choise of the parton distribution functions (PDFs) and variations of 
% $\mu$ from  $0.5\times M_LQ$ to  $2 \times M_LQ$ added in quadrature  
%
\begin{table}[ht]
\begin{center}
\caption{$LQ\overline{LQ} \rightarrow \nu\bar{\nu}jj$ analysis.
Number of data, background and signal events after all selections for different LQ signals.}
\begin{tabular}{|c|c|c|c|c|}
\hline $M_{LQ}$ & $(\etmiss,H_T) $& Data & Background & Signal    \\
       GeV        &  GeV         &       &            &           \\
\hline
       140  &  (75,150)  &  352 & 328$\pm$ 11$^{+56}_{-57}$  & 229$\pm$8$^{+24}_{-23}$\\
       200  &  (125,300) &  12  & 10.6$\pm$1.7$^{+4.0}_{-2.0}$ & 13.7$\pm$0.6$^{+1.8}_{-2.0}$ \\
\hline 
\end{tabular}
\label{tab:lq_allgen}
\end{center}
\end{table}
\section{Search for pair production of first generation leptoquarks}
A search for pair production of first generation leptoquarks was 
performed with 1~fb$^{-1}$ of data on the final states with two electrons 
and two jets ($LQ\overline{LQ} \rightarrow eejj,~\beta = 1$), 
or one electron, two jets and $\etmiss$ ($LQ\overline{LQ} \rightarrow e\nu jj,~\beta = 0.5$).
Data samples were collected with combinations of single electron and electron plus jet
triggers. 

In the $eejj$ analysis the selected events were required to have 
at least two isolated electrons with $p_T > 25$~GeV (one of the electrons
should be detected in $|\eta| < 1.1 $)  and 
at least two jets with $p_T > 25$~GeV in the $|\eta| < 2.5$ region. 
The dielectron invariant mass $M_{ee}$ and the transverse scalar energy $S_T$ 
(the scalar sum of the momenta of the two electrons and the two highest $E_T$ jets) 
were used as discriminant variables against the background processes 
dominated by $Z/\gamma \rightarrow e^{+}e^{-}$ events. 
The best sensitivities to leptoquark signals 
(with (20--23)\% acceptances for LQ masses in 250--300~GeV range)
were obtained with $S_T > 400$~GeV and $M_{ee} > 110$~GeV. 
No data events remained after all selections compared to the 
estimated backround of $1.51\pm0.12(stat)\pm0.04(syst)$ events.
  
In the $e\nu jj$ analysis events were required to have $\etmiss > 35$~GeV, 
one isolated electron with $p^{e}_T > 25$~GeV in $|\eta| < 1.1$, and at 
least two jets in the $|\eta| < 2.5$ region with $p_T > 40$~GeV (leading jet) 
and $p_T~>~25$~GeV (second leading jet). A veto on a second electron 
was applied to avoid overlap with the $eejj$ selections. 
Cuts on $\etmiss$, $p^{e}_T$, $M_T(e,\etmiss)$ (the transverse invariant mass of the electron and $\etmiss$) 
were optimized against the best expected 95\%~C.L. limit on signal cross section. 
After requiring $\etmiss > 80$~GeV, $p^e_T~>~80$~GeV, $M_T(e,\etmiss) > 130$~GeV eight events 
remained in data, while the background
estimation gave $9.8\pm0.8\pm0.8$. The dominant background contributions are from 
$W(\rightarrow l\nu) + jets$ and $t\bar{t}$ processes. 
Signal acceptances varied between (18.5--20)\% for LQ mass in the range 250--300~GeV.
 
\begin{figure*}[ht]
\centerline{
   \includegraphics[scale=0.40]{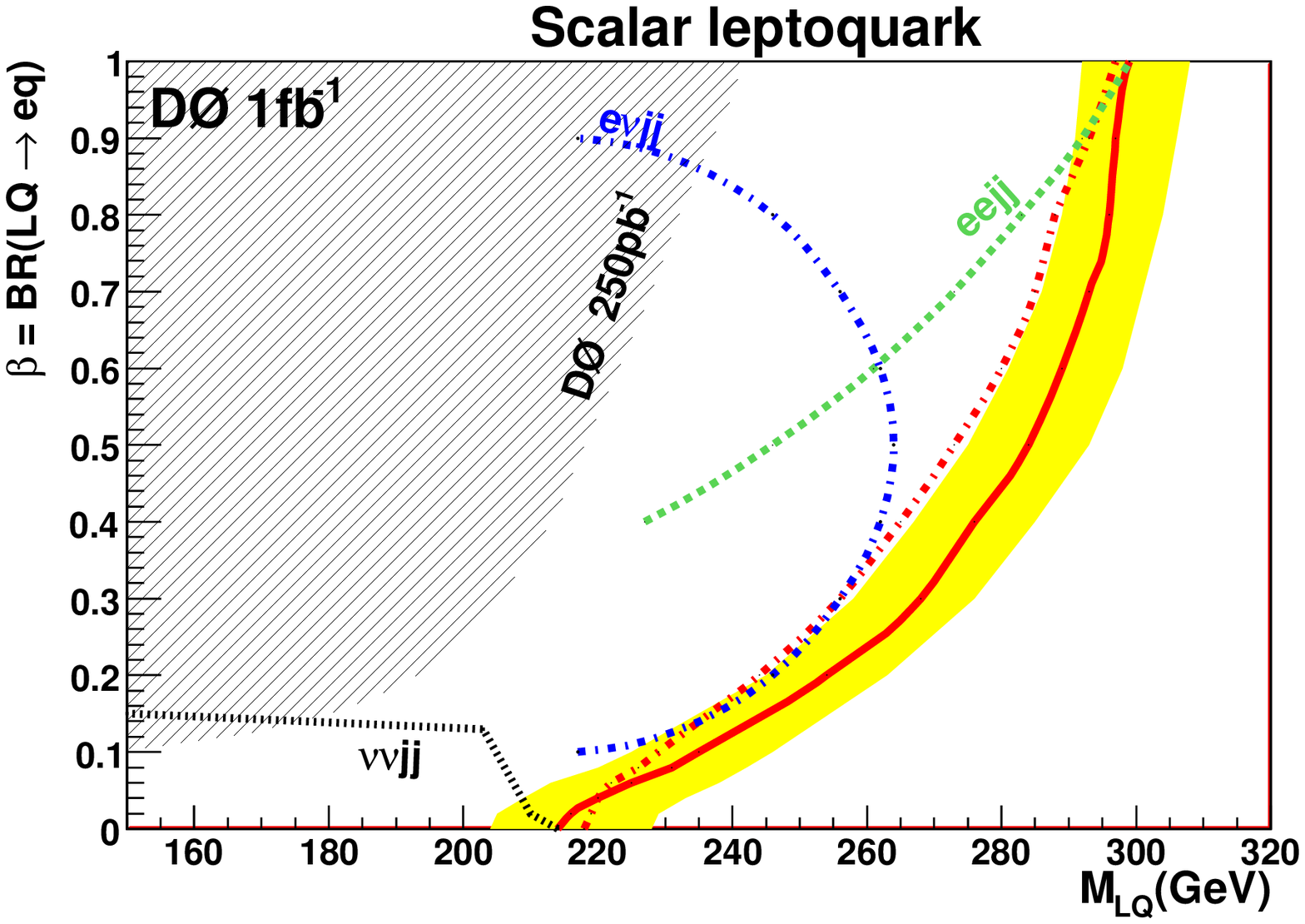}
   \includegraphics[scale=0.40]{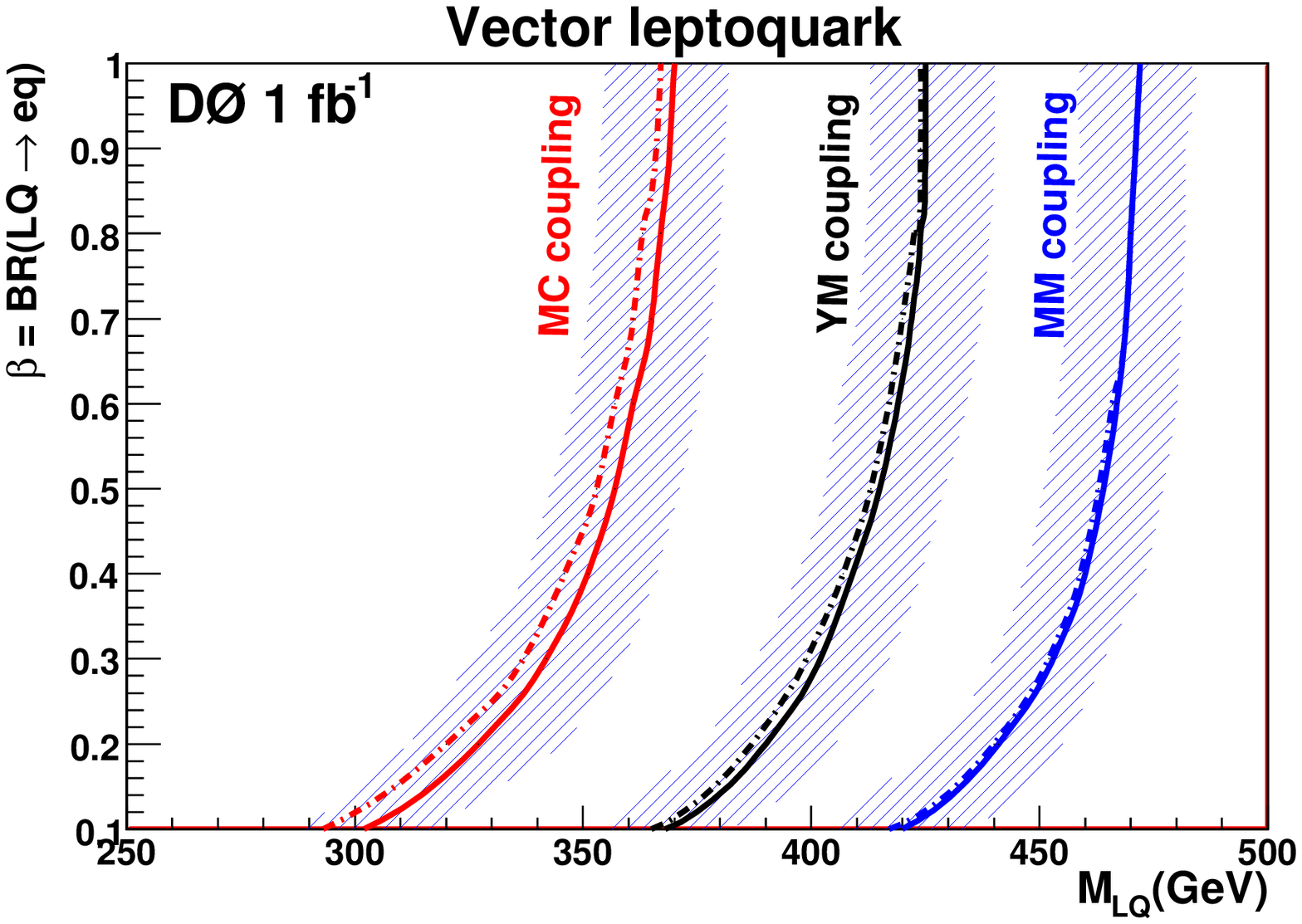}
}
 \leftline{ \hspace{2cm}{\bf (a)} \hfill\hspace{2cm} {\bf (b)} \hfill}
 \caption{\label{fig:lq1}
% The observed (solid lines) and expected (dot-dashed lines) 95\%~C.L. mass limits in the $(M_{LQ},\beta)$ plane for
% a) first generation scalar leptoquarks and b)  first generation vector leptoquarks, shown for different LQ-gluon couplings.
% The shaded bands correspond 
a) The observed (red solid line) and expected (red dot-dashed line) 95\%~C.L. first generation scalar leptoquarks 
mass limits in the $(M_{LQ},\beta)$ plane. The yellow band corresponds to the uncertainties due to the choice of the PDFs and variations of the renormalization scale on the observed limits. Also shown are the observed limits individually obtained
with the 1~fb$^{-1}$ $eejj$ and $e\nu jj$ and 2.5~fb$^{-1}$ $\nu\nu jj$ analysis. The hatched region was excluded by the 
previous D0 search with a 250~pb$^{-1}$ dataset.
b) The observed (solid lines) and expected (dot-dashed lines) 95\%~C.L. first generation vector leptoquark 
mass limits in the $(M_{LQ},\beta)$ plane shown for different $LQ-gluon$ couplings. Regions to the left of the curves 
are excluded. The hatched bands correspond to the uncertainties due to the choice of the PDFs and variations 
of the renormalization scale on the observed limits.}
\end{figure*}

For both the $eejj$ and $e\nu jj$ channels no deviation from the SM predictions were found 
and the lower 95\%~C.L. limits on a scalar LQ mass of 299~GeV and 284~GeV respectively were set. 
These two analysis were combined with the 2.5~fb$^{-1}$ $\nu\nu jj$ search,
and limits on LQ mass was determined for  different values of $\beta$. 
Figure~\ref{fig:lq1}(a) shows the observed and expected LQ mass limits 
in the $(\beta,M_{LQ})$ mass plane. 

Vector leptoquarks mass limits were obtained using the leading order 
theoretical production cross section with the renormalization 
scale $\mu = M_{LQ}$. Acceptances for vector leptoquarks are similar 
to thouse for scalar leptoquarks with the same $M_{LQ}$ and the same selection as in scalar $eejj$ and
$e\nu jj$ analyses were applied. Figure~\ref{fig:lq1}(b) shows combined limits on vector leptoquark mass
for different couplings in the $(\beta,M_{LQ})$ mass plane. For $\beta = 0.5$ 
the 95\% C.L. lower limits on first generation vector leptoquarks masses are correspondingly
357~GeV, 415~GeV and 464~GeV for the Minimal, Yang-Mills and Minus Minus couplings. 
%A cut was required for events with $>=$ jets  in the $\Delta\phi(e,etmiss)-\etmiss$ plane against the events with fake electron due to jet misidentification and with mismeasured $\etmiss$. 
%
\section{Search for pair production of second generation scalar leptoquarks}
A search for $\mu\mu jj$ and $\mu\nu jj$ final states was made with 1~fb$^{-1}$ data sample.
The events were collected with a combination of single muon triggers.  

In the $\mu\mu jj$ analysis events were required to have at least two muons with $p_T > 20$~GeV and
at least two jets with $p_T > 25$~GeV. The invariant mass $M(\mu,\mu)$, 
reconstructed from the two muons of highest
$p_T$  and the $S_T$ (the sum of two leading jets and the two leading
muons transverse momenta) were required to be greater 100~GeV and 200~GeV, respectively. 
No excess of data over the SM backgrounds (dominated by $Z(\rightarrow \mu\mu) + jets$ events) 
was observed after these selections. The multijet background in this final state is negligible.
The $M(\mu,\mu)$, $S_T$ and four muon-jet invariant masses $M({\mu_i},jet_{i})$ 
were used in a neural net (NN) trained separately for the each of analyzed LQ signals and SM background samples. 
%The shape of the neural net output variable distribution was used to optimize sensitivity 
The bin edges of the NN output discriminant were chosen to concentrate
background events into a few bins while the remaining bins have a significant signal to background ratio.
All bins were then treated as individual channels in the calculation of the cross section limits. 
This method allows optimize signal sensitivity using the shape of the NN discriminant without cutting on it. 

The $\mu\nu jj$ decay channel gives a $\mu \etmiss jj$ signature. 
The main background are $W ( \rightarrow \mu\nu ) + jets$
production and multijet events in which a jet is misidentified as an isolated muon. 
Events were required to have $\etmiss > 30$~GeV, exactly one muon with $p_T > 20$~GeV,
and at least two jets with $p_T > 25$~GeV. 
To suppress multijet backrounds and remove events with mismeasured muon $p_T$ 
the transverse mass $M_T(\mu, \etmiss)= \sqrt{2p_{T}(\mu\mu)\etmiss (1 -cos(\Delta\phi(\mu, \etmiss))}$
was required to be greater then 110~GeV and the azimutal angle between the $\etmiss$ and the muon 
was constrained to be smaller then 3.0 radians. 
The $S_T=p^{\mu}_t + p^{jet1}_T + p^{jet2}_T + \etmiss$ was required to be greater 200~GeV.
Data were found to agree with the estimated background. 
The $M_T(\mu,\etmiss)$, $S_T$, two transverse masses $M(\etmiss,jet_i)$, 
and two invariant masses $M(\mu,jet_{i})$ were used 
to calculate a neural net discriminant for each analyzed leptoquark mass and for different $\beta$ values. 

\begin{figure*}[ht]
\centerline{
   \includegraphics[height=67mm,width=80mm]{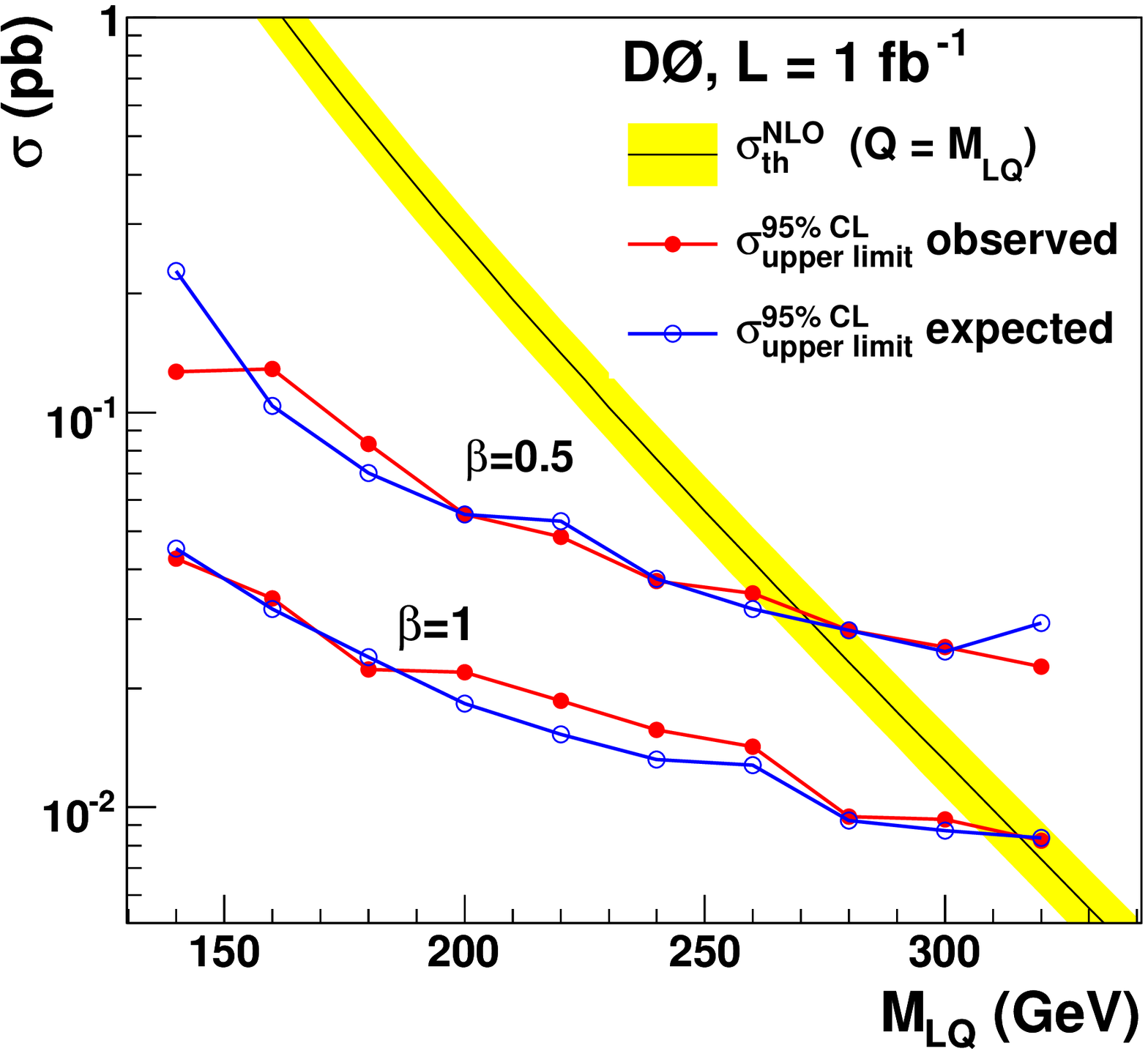}
   \includegraphics[scale=0.40]{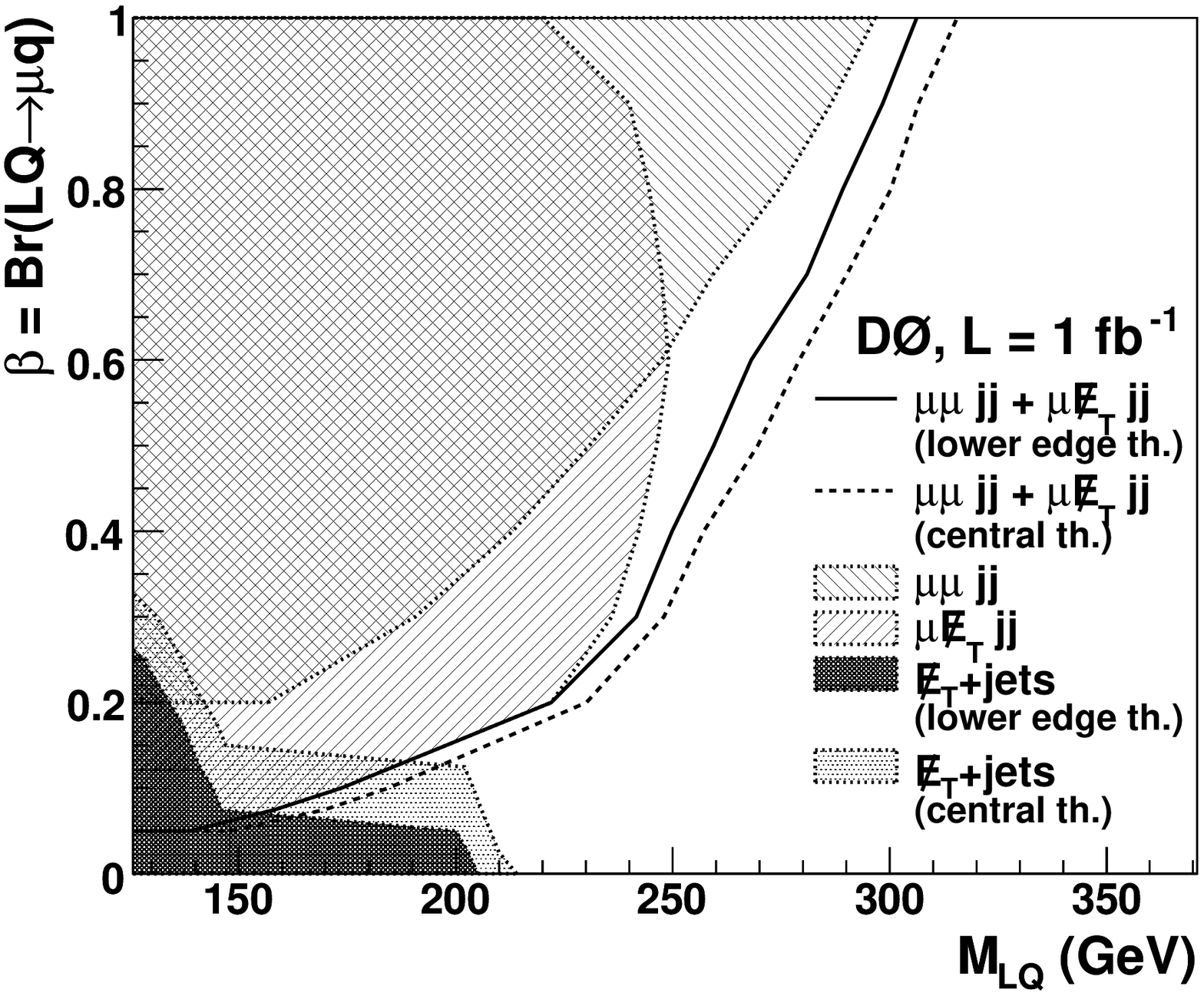}
}
 \leftline{ \hspace{2cm}{\bf (a)} \hfill\hspace{2cm} {\bf (b)} \hfill}
 \caption{\label{fig:lq2}
a) The observed (red lines) and expected (blue lines) 95\%~C.L. cross section limits for second generation scalar leptoquark pair production for $\mu\mu jj$ and $\mu\nu jj$ final states. The nominal NLO production cross section
is shown with shaded bands corresponding to the uncertainties due to choice of PDFs and variations 
the factorization and renormalization scale.
b) The observed 95\%~C.L. excluded regions in the $(M_{LQ},\beta)$ plane for $\mu\mu jj$ and $\mu\nu jj$. 
The region to the left of the solid line is excluded by the combination of these analyses obtained using the lower bound of NLO theory for which variation of the renormalization scale $\mu = + 2M_{LQ} $ and the PDF errors were used. The dashed
line corresponds to the exclusion with the nominal NLO predictions. At small $\beta$ the exclusion region based
on 2.5~fb$^{-1}$  $\nu\bar{\nu}jj$ search is also shown.}
\end{figure*}

Figure~\ref{fig:lq2}(a) shows the observed and expected 95\%~C.L. limits 
for second generations scalar leptoquark pair production cross section 
for $\beta = 1$, and $\beta = 0.5$ as functions of $M_{LQ}$.  
The observed 95\%~C.L. exclusion regions in the $(M_{LQ},\beta)$ plane for $\mu\mu jj$ 
and $\mu \etmiss jj$ selections and their combinations are shown in Figure~\ref{fig:lq2}(b). 
For  $\beta = 1$,  $\beta = 0.5$ and $ \beta = 0.1$ combined analyses excluded second generation 
leptoquarks with masses $M_{LQ} < 316$~GeV, $M_{LQ} < 270$~GeV and $M_{LQ} < 185$~GeV (using the lower bound of the NLO theory) 
\section{Searches for pair production of third generation scalar leptoquarks}
Third generation leptoquarks would decay into a $b$ or $t$ quarks and to a $\tau$ or tau neutrino
depending on their electric charge. The $\tau b$ decay is possible for $LQ$ with charge 2/3 or 4/3 (section~\ref{tau_b})
and charge 1/3 $LQ$ will decay to $\nu_{\tau}\bar{b}$ quark (section~\ref{nu_b}). The $t\tau$ and $t\nu_{tau}$ decays
are suppressed due to large top quark mass. Both analysis used a neural net $b$-tagging tool 
based on the DO tracker information to increase the signal sensitivity.
\subsection{Search for $LQ\overline{LQ} \rightarrow \tau\tau b\bar{b}$}
\label{tau_b}
A search was conducted for $\tau^{+}\tau^{-}b\bar{b}$ final state
where one of the $\tau$ decays hadronically and the other through $\tau \rightarrow \mu\nu\nu$. 
Thus the event signature is two $b$-jets, an isolated  muon, a $\tau$ candidate,
and missing transverse energy.  
A 1~fb$^{-1}$ dataset was collected  with a set of triggers required either a 
single muon or a muon in association with jets. 

\begin{figure}[ht]
\centering
\includegraphics[width=80mm,height=67mm]{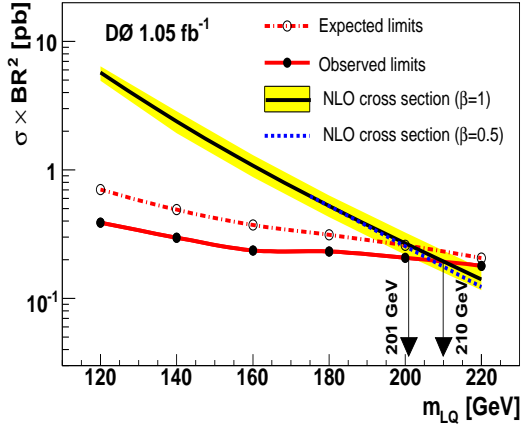}
\caption{Observed (solid line) and expected (dash-dotted line) 95\% C.L. limits on pair production of third generation leptoquarks decaying to $\tau b$. The yellow shaded band shows the uncertainty on the theoretical prediction. The dashed line indicates the threshold effect for the $t\nu_{\tau}$ channel.} 
\label{fig:lq3_taub}
\end{figure}

Neural networks were formed to identify hadronic tau candidates ($\tau_h$)
with calorimeter and track information for each of three possible decay modes:
$\tau^{\pm} \rightarrow \pi^{\pm}\nu$, 
$\tau^{\pm} \rightarrow \pi^{\pm}\pi^{0}_{s}\nu$, and 
$\tau^{\pm} \rightarrow \pi^{\pm}\pi^{\pm}\pi{\mp}\pi^{0}_{s}\nu$.  
Events were required to have exactly one isolated  muon with $p_T > 15$~GeV, 
at least two jets with $p_T > 20$~GeV, a $\tau_h$ candidate with $p_T > $15--20GeV.
and no electrons with $p_T > 12$~GeV.  After these selections the estimated backround is dominated by
multijet QCD events, events from $W/Z+light\ jets$ processes, and from $t\bar{t}$ production.
To increase sensitivity to LQ signal the $m^{*}$ parameter defined as 
$m^{*}= \sqrt{2E^{\mu} E^{\nu}(1-cos\Delta\phi(\etmiss,\mu))}$, $E^{\nu}=\etmiss(E^{\mu}/p^{\mu}_T)$  was required 
to be less then 60~GeV. Remaining events were divided in two subsamples with one or $\geq$ 2 $b$-tagged jets.
Table~\ref{tab:lq3_taub} shows the number of data, backround and signal ($M_{LQ} = 200$~GeV) events after different selections.
No excess above the expected background were observed. Limits on LQ productions cross section were set using 
distributions of the $S_T=p^{\mu}_T+p^{\tau}_T + p^{jet_1}_T + p^{jet_1}_T$ parameter which is in average higher for
the signal events. Figure~\ref{fig:lq3_taub} shows  95\% C.L. cross section limits as a function of the LQ mass. 
The obtained lower limit on third generation leptoquark mass of 210~GeV 
(207~GeV if $t\nu_{\tau}$ decay is possible) 
is the most stringent for this decay mode to date. 
\begin{table}[ht]
\begin{center}
\caption{$LQ\overline{LQ} \rightarrow \tau\tau b\bar{b}$ analysis.
Number of data, background and signal ($M_{LQ} = 200$~GeV) events after different selections.}
\begin{tabular}{|c|c|c|c|c|}
\hline Source     &  $m^{*}<60$~GeV & 1 $b$-tag  & $\geq$ 2 $b$-tags    \\
\hline          
 Data             &   94          &          15   &   1          \\
 Backround        & 109.2$\pm$5.7 & 19.6$\pm$2.5  & 4.8$\pm$0.1  \\
 LQ signal        &   7.4$\pm$0.1 & 3.4$\pm$0.1   & 2.6$\pm$0.1  \\
\hline 
\end{tabular}
\label{tab:lq3_taub}
\end{center}
\end{table}
\subsection{Search for third generation leptoquark pairs in acoplanar $b$-jet events}
\label{nu_b}
A search was made for $LQ\overline{LQ} \rightarrow \nu\bar{\nu}  b\bar{b}$.
The corresponding detector signature is two acoplanar $b$-jets 
from the $b$-quarks and the missing energy due to escaping neutrinos. 
The data sample was collected using jet plus missing energy triggers 
and corresponds to an integrated luminosity of 4~fb$^{-1}$. 

Events were required to have $\etmiss > 40$~GeV, exactly two or three jets of $p_T > 20$~GeV ($p_T > 50$~GeV for
the leading jet). The two leading jets had to be acoplanar ($\Delta \phi(jet_1,jet_2) < 165^{\circ}$), $b$-tagged, and have
the energy fraction in the event $(E_{T}^{jet1} +  E_{T}^{jet2} ) / (\Sigma_{jets} E_T)$  greater
then 0.9. To reduce the contribution from $W\rightarrow l\nu$ decays, events with isolated electrons or isolated 
muons with $p_T > 15$~GeV were vetoed. The QCD multijet backround was suppressed with removal of events
where the $\etmiss$ direction overlapped a jet in $\phi$ and events where the direction of $\etmiss$ is not 
aligned with the missing track $\ptmiss$ (the negative of the vectorial sum of 
charged particles transverse momenta). 
Similar to the generation independent search, cuts on $\etmiss$ and $H_{T}$ were optimized
for the LQ signals of different masses. 

\begin{figure}[ht]
\centering
\includegraphics[width=80mm]{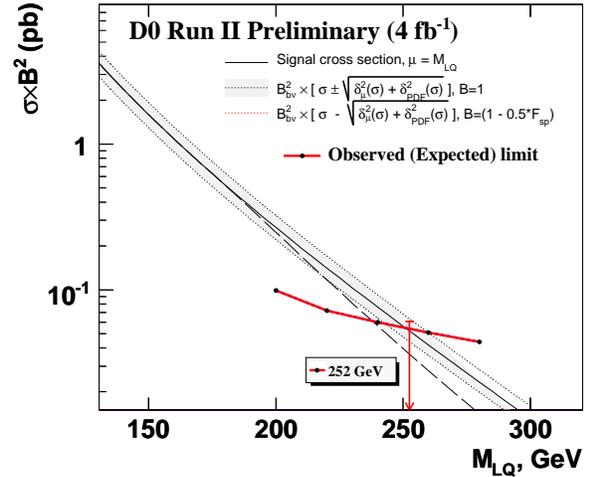}
\caption{The 95\%~C.L. limit on  $\sigma \times B^2_{b\nu}$ (points plus
 solid line) as a function of $M_{LQ}$ for the pair production 
 of third generation leptoquarks. The expected cross section limits are the same as the observed. 
 The grey shaded band shows the PDF and the renormalization scale error bounds on the NLO
 production cross section (solid line) calculated for $\mu = M_{LQ}$. The long dashed line shows $\sigma*B^{2}(LQ \rightarrow \nu b)$ for the $B(\nu b)$ = $B(t\tau)= 0.5$ times the phase suppression factor for the $t\tau$ channel.} 
\label{fig:lq3_nub}
\end{figure}

Table~\ref{tab:lq3_nub} shows the number of data, 
estimated background (dominated by the events from top quark production and from $W/Z + b\bar{b}/c\bar{c}\ jets$,
contribution of QCD multijet backround is negligible) and LQ events after all selections. 
No signal was observed and the 95\%~C.L.~limits
on charge 1/3 scalar third generation leptoquark production cross section were set (shown in Fig~\ref{fig:lq3_nub}). 
Assuming $B(LQ\rightarrow \nu b) = 1$ a mass limit of 252~GeV was obtained. 
If $t\tau$ decay is possible then the limit is $M_{LQ} > 239$~GeV.
\begin{table}[ht]
\begin{center}
\caption{$LQ\overline{LQ} \rightarrow \nu\bar{\nu} b\bar{b}$ analysis.
Number of data, background and signal events after all selections for different LQ signals.}
\begin{tabular}{|c|c|c|c|c|}
\hline $M_{LQ}$ & $(\etmiss, H_T)$   & Data & Background &     Signal      \\
       GeV     &   GeV             &      &            &    (acpt, \%)   \\
\hline
       200  &  (130,220) &  7 &  7.1$\pm$0.5$\pm$1.2  & 23.2$\pm$0.8$\pm$3.3 (2.1)\\
       280  &  (150,240) &  3 &  3.2$\pm$0.3$\pm$0.6  & 3.9$\pm$0.1$\pm$0.5 (3.4) \\
\hline 
\end{tabular}
\label{tab:lq3_nub}
\end{center}
\end{table}
\section{Summary}
Searches for pair production of leptoquarks of all three generations 
were performed in 1--4~fb$^{-1}$ D0 data samples. 
All presented analysis are in good agreement with the SM predictions. 
No leptoquarks signals were observed, and a set of 95\% C.L. limits on
LQ masses have been obtained improving previous Tevatron results.
More details on the presented analyses can be found at~\cite{d0_np}.
%%%%%%%%%%%%%%%%%%%%%%%%%%%%%%%%%%
% \begin{acknowledgments}
% This document is adapted from the instructions provided to the authors
% of the proceedings papers at CHARM~07, Ithaca, NY~\cite{charm07},  
% and from eConf templates~\cite{templates-ref}.
% \end{acknowledgments}

\bigskip % extra skip inserted
% Create the reference section using BibTeX:
%\bibliography{basename of .bib file}

\begin{thebibliography}{9}   % Use for  1-9  references
%\begin{thebibliography}{99} % Use for 10-99 references
\bibitem{lq_theory} J.C.~Paty, A.~Salam, Phys.\ Rev.\ D  {\bf10}, 275 (1974);
                    H.~Georgi, S.~Glashow, Phys.\ Rev.\ Lett. {\bf32}, 438 (1974);
                    B.~Schrempp, F.~Schrempp, Phys.\ Lett.\ B {\bf 153}, 101 (1985). 

\bibitem{d0_upgrade} D0 Collaboration, V.~Abazov {\it et al.},
Nucl.\ Instrum.\ Methods\ A {\bf565}, 463 (2006).

\bibitem{d0_np} http://www-d0.fnal.gov/Run2Physics/WWW/results/np.htm
%
\end{thebibliography}

%
\end{document}